\documentclass[prb,showpacs,twocolumn]{revtex4}
\newcommand {\Neff}{$N_{\mathrm{eff}}$}

\newcommand {\EparaC}{{\bf E} $\parallel c$}
\newcommand {\EparaA}{{\bf E} $\parallel a$}

\usepackage{graphicx}
\usepackage{bm}
\usepackage{pifont}
\usepackage{amsmath}
\begin{document}
\title{Pseudogap-related charge dynamics in layered-nickelate \textit{R}$_{2-x}$Sr$_x$NiO$_4$ ($x\sim 1$)}
\author{M. Uchida$^1$, Y. Yamasaki$^2$, Y. Kaneko$^3$, K. Ishizaka$^1$, J. Okamoto$^2$, H. Nakao$^{2,4}$, Y. Murakami$^2$, and Y. Tokura$^{1,3,5}$} 
\affiliation{
$^1$ Department of Applied Physics, University of Tokyo, Tokyo 113-8656, Japan \\
$^2$ Photon Factory and Condensed Matter Research Center, Institute of Materials Structure Science, High Energy Accelerator Research Organization, Tsukuba, Ibaraki 305-0801, Japan \\
$^3$ Multiferroics Project, ERATO, Japan Science and Technology Agency (JST), Tokyo 113-8656, Japan \\
$^4$ CREST, Japan Science and Technology Agency (JST), Tokyo 102-0076, Japan \\
$^5$ Cross-Correlated Materials Research Group (CMRG) and Correlated Electron Research Group (CERG), RIKEN Advanced Science Institute (ASI), Wako 351-0198, Japan
}
\date{\today}
\begin{abstract}
Charge dynamics and its critical behavior are investigated
near the metal-insulator transition of layered-nickelate $R_{2-x}$Sr$_x$NiO$_4$ ($R=\mathrm{Nd}$, Eu).
The polarized x-ray absorption spectroscopy experiment clearly shows the multi-orbital nature
which enables the $x^2-y^2$-orbital-based checkerboard-type charge ordering or correlation to persist up to the critical doping region ($x\sim 1$).
In the barely metallic region proximate to the charge-ordered insulating phase, 
the nominal carrier density estimated from the Hall coefficient markedly decreases in accord with development of the pseudogap structure in the optical conductivity spectrum,
while the effective mass is least enhanced.
The present findings combined with the results of recent angle-resolved photoemission spectroscopy show that
the pseudogap in the metal-insulator critical state evolves due to the checkerboard-type charge correlation
to extinguish the coherent-motion carriers in a characteristic momentum (\textbf{k})-dependent manner with lowering temperature.
\end{abstract}
\pacs{71.30.+h, 78.20.-e, 74.72.Kf, 78.70.Dm}
\maketitle

\section{Introduction}

Charge ordering is one of the generic phenomena in strongly correlated electron systems \cite{MIT}
and occurs with a variety of ordering patterns (e.g. in forms of stripe and checkerboard)
depending on the band filling of the conduction electrons.
The ordering in the charge sector by strong correlation is occasionally accompanied with
the concomitant or subsequent orders of spin, orbital, and lattice degrees of freedom,
which enrich the relevant electronic phase transitions.
One such example is melting of the charge order in the critical states, which leads to emergence of,
for example, the high-temperature superconductivity in layered cuprates \cite{super}
and the ferromagnetic metal in colossal magnetoresistive manganites \cite{CMR}.
In the metallic state derived from the melting of long-range orders,
the charge as well as relevant spin/orbital correlations may have a large impact on the charge dynamics,
giving rise to rich spatiotemporal charge responses such as formation of pseudogap structure in the momentum (\textbf{k})-space.
The purpose of this work is to establish the link between the spectroscopically-resolved pseudogap and the charge transport properties
in barely metallic layered nickelates, a typical electron-correlated system with charge-ordering instability.

\begin{figure}
\begin{center}
\includegraphics*[width=7.5cm]{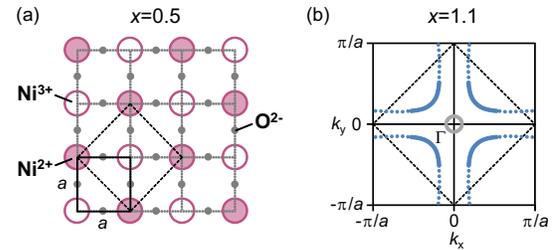}
\caption{(Color online).
Schematic of electronic structure evolution with hole doping in the single-layer nickelate $R_{2-x}$Sr$_x$NiO$_4$.
(a) Checkerboard (CB)-type charge ordering on the NiO$_2$ plane at $x=0.5$.
(b) Underlying two-dimensional hole Fermi surface with \textbf{k}-dependent pseudogap at $x=1.1$,
reproduced from the previous ARPES study. \cite{NiFS1}
Solid and dotted curves represent the Fermi momentum positions
where one can see a quasiparticle peak and a high-energy pseudogap structure, respectively.
A three-dimensional electron pocket centered only in $k_z=0$ plane is also shown.
}
\label{fig1}
\end{center}
\end{figure}

Layered-nickelate $R_{2-x}$Sr$_x$NiO$_4$ ($R$: rare earth elements) is a prototypical system showing
both the charge ordering \cite{MIT} and metal-insulator transition with hole doping. \cite{Nipoly1, Nipoly2}
Besides the diagonal stripe-type commensurate or incommensurate charge-spin ordering around $x=1/3$,
\cite{Nistripeele, Nineutrontra, Nisigma2, Nistripedoping, Niorthostripe, NiCIC}
the checkerboard (CB)-type charge ordering as shown in Fig. 1(a) emerges on the NiO$_2$ plane
centered around the other fixed point $x=1/2$. \cite{NisigmaCB, Nineutron, Nisigma1, NioptCB}
At $x=0.5$, the $x^2-y^2$-orbital-based CB-type charge order develops below $T_{\mathrm{CO}} \sim 500$ K
under the strong intersite Coulomb repulsion and cooperative Jahn-Teller distortion.
The CB-type charge ordered phase has been actually confirmed at least up to $x=0.7$, while it has never been observed below $x=0.5$.
In the in-plane optical conductivity spectra at the ground state, a real-gap structure exists with a maximum gap size at $x=0.5$, \cite{Nisigma1}
and it gradually evolves into a pseudogap in the metallic region proximate to the insulating phase. \cite{Nithin}
Recently the corresponding structure has been found in a \textbf{k}-dependent form 
by angle-resolved photoemission spectroscopy (ARPES) at $x=1.1$. \cite{NiFS1}
As sketched in Fig. 1(b), the main Fermi surface is a ($\pi, \pi$)-centered large cylindrical hole surface,
while the electron-pocket-like Fermi surface with dominant $3z^2-r^2$ orbital character is formed around the $\Gamma$ point. \cite{NiFS2}
The gap of about 0.1 eV is observed to open around the ($\pi, 0$) region at low temperature.
This feature bears analogy to the high-energy pseudogap observed 
in underdoped superconducting cuprates \cite{PG, Cureview} and also in colossal magnetoresistive manganites \cite{pgMn}.
Since the antiferromagnetic interaction in the nickelate parent material is one order of magnitude smaller than that of the cuprates, \cite{NiJ}
it has been suggested that a plausible origin of the pseudogap is not the spin correlation but rather the remnant charge correlation. \cite{NiFS1}
Here first we discuss the doping variation of the hole orbital states based on analysis of linearly polarized x-ray absorption spectra (XAS),
and clarify the orbital nature of the CB-type charge ordering or correlation, which can persist up to the metal-insulator critical doping region.
Then we report on temperature and doping variations of the high-energy pseudogap as probed by optical spectroscopy
and their interrelation with anomalous charge transport in the course of the metal-insulator transition.
The present systematic study in the nickelate system can give important hints
to understanding of generic charge dynamics under the influence of strong real-space charge correlation.

\section{Experiment}

Single-crystalline samples of La$_{2-x}$Sr$_x$NiO$_4$ ($x=0.5$), Nd$_{2-x}$Sr$_x$NiO$_4$ ($x=0.3$-1.3),
and Eu$_{2-x}$Sr$_x$NiO$_4$ ($x=0.7$-1.2) were grown by a floating-zone method.
The sample growth above $x=0.8$ was carried out in a high-pressure oxygen atmosphere up to $p_{\mathrm{O}_2}=60$ atm.
$R=\mathrm{Nd}$ and Eu crystals obtained in the highly-doped region were adopted for the following measurements;
in this system the electronic properties have been reported to be
least affected by variation of the $R$ species. \cite{NiFS1, Nisigma1, Nineutron, Niorthostripe}

XAS experiments were carried out at undulator beam line BL-16A of KEK Photon Factory.
The O $K$-edge data were taken in the fluorescence-yield (FY) mode with a typical attenuation length of about 600 \AA.
The measurements were performed in high vacuum of $1 \times 10^{-8}$ torr at 50 K.
The XAS spectra for {\EparaA} and {\EparaC} were measured
utilizing vertically and horizontally polarized soft-x-rays in grazing incidence geometry to the $ab$ plane, respectively.
The incidence angle relative to the surface was set to $10^{\circ}$ and the degree of the linear polarization was about 95\%.
Range of the resultant error is shown including one caused in the analysis procedure.

Longitudinal and Hall resistivities and heat capacity were measured with the Quantum Design Physical Property Measurement System.
In-plane reflectivity spectra in the temperature range of 10-600 K were measured between 0.01 and 5 eV.
The measurements above room temperature were performed in a flow of O$_2$ gas to prevent the reduction,
and we confirmed that the degradation at elevated temperature was negligible by thermally-repeated measurements.
Synchrotron radiation at UV-SOR, Institute for Molecular Science, was utilized for the measurements (only at room temperature) between 5 and 40 eV.
For Kramers-Kronig analysis, the spectrum above 40 eV was extrapolated by a $\omega ^{-4}$ function,
while below 0.01 eV the Hagen-Rubens relation or the constant reflectivity was assumed depending on the metallic or insulating ground state.

\section{Results and discussion}

In Fig. 2(a) we show the polarization-dependent O $K$-edge spectra at $x=0.5$ for {\EparaA} and {\EparaC}.
Aside from the main peak ($\mathrm{O1}s \to \mathrm{O2}p$) at about 536 eV,
some pre-edge peaks assigned to excitations brought about by the $p-d$ hybridization ($\mathrm{O1}s \to \mathrm{Ni3}d$)  \cite{NdpolyOKNiL, LasingpolyOK}
are clearly observed in the lower energy region (3 peaks for {\EparaA} and 2 peaks for {\EparaC}).
Considering all the possible transition processes as sketched in Fig. 2(b),
these pre-edge peaks at about 532 and 529 eV for $x=0.5$ are ascribable to the processes A and B, as assigned in the previous work. \cite{LasingOK}
The two initial states are Ni$^{2+}$ one and Ni$^{3+}$ one with unoccupied $x^2-y^2$ orbital, respectively,
indicating that the holes below $x=0.5$ are certainly doped into the $x^2-y^2$ orbital.
Figures 2(c) and 2(d) show the detailed doping variation of the O $K$-edge spectra in the pre-edge region for {\EparaA} and {\EparaC}, respectively.
Judging from the change below and above $x=0.5$ for {\EparaC},
an additional peak, which is clearly discerned above $x=0.6$, can be assigned to the process C,
indicating the existence of Ni$^{3+}$ site with unoccupied $3z^2-r^2$ orbital.
Above $x=1.0$ the total peak positions largely shift to lower energy both for {\EparaA} and {\EparaC};
this may arise from the appearance of an additional peak, which can be assigned to the process D relevant to the high-doping induced Ni$^{4+}$ sites.

\begin{figure}
\begin{center}
\includegraphics*[width=8.6cm]{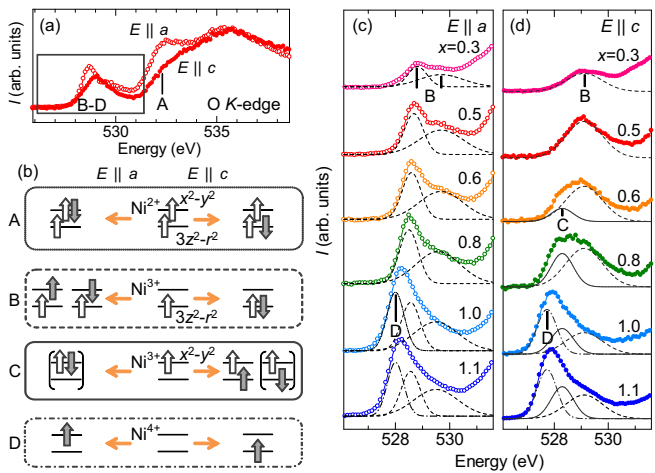}
\caption{(Color online).
(a) Polarization-dependent XAS spectra at O $K$-edge for {\EparaA} and {\EparaC}.
Some pre-edge peaks as labeled with letters (A-D) are clearly discerned in the lower energy region (528-533 eV) next to the main peak centered at about 536 eV.
(b) Schematic of all the transition processes corresponding to the pre-edge peaks.
Filled arrows added in the left and right sides represent the electrons which are transferred from the O 1$s$ state to the initial Ni 3$d$ $e_g$ orbital state shown in the middle.
Detailed doping variation of the O $K$-pre-edge spectra for (c) {\EparaA} and (d) {\EparaC}.
Gaussian fittings for the pre-edge peaks corresponding to the transition processes (B, C, and D) as depicted in (b) are indicated by dashed, solid, and chain curves, respectively.
The peaks corresponding to the bracketed final states in the panel C of (b) cannot be clearly discerned
probably because such higher-energy peaks are easily hidden by the main or other pre-edge peaks.
We confirm that variation of the fitting procedures even considering the extra peaks causes
negligible changes in the estimated site occupancy shown in Fig. 3(a) within the error bar and no difference in our conclusion.
}
\label{fig2}
\end{center}
\end{figure}

By fitting the pre-edge peaks B-D with Gaussian curves and comparing the intensity variations for {\EparaC}, 
the valence- and orbital-state changes of the Ni site with doping $x$ can be roughly estimated.
Figure 3(a) summarizes doping variation of the estimated site occupancy of the respective valence and orbital states.
Up to $x=0.5$, the Ni$^{3+}$ site with unoccupied $x^2-y^2$ orbital increases nearly linearly in density.
This change corresponds to the fact that the $x^2-y^2$-orbital electrons and holes form the stripe- and CB-type charge orders on the NiO$_2$ plane.
Above $x=0.5$, on the other hand, the Ni$^{3+}$ site with the $3z^2-r^2$ hole starts to increase, while that with the $x^2-y^2$ hole is nearly constant, namely 1/2 per Ni site.
This nonmonotonic variation strongly indicates that the excess holes above $x=0.5$ are mainly doped into the $3z^2-r^2$ orbital
and thus the $x^2-y^2$-orbital-based CB-type charge ordering or correlation robustly persist far beyond the fixed point $x=0.5$.
A similar $x$-independent commensurate order pattern has been reported for a wide doping region $0.3\le x\le 0.5$
in the so-called CE-type spin-charge-orbital ordering of Pr$_{1-x}$Ca$_x$MnO$_3$.
This has been attributed to occupancy of the other $3z^2-r^2$ orbital states ($z\parallel c\mathrm{-axis}$) by the extra electrons,
which has been confirmed by the electron diffraction measurement. \cite{Asaka}
Thus, the present behavior in the layered nickelate can be regarded as the hole counterpart to the manganite case, as also expected in the previous work. \cite{Nisigma1}
The observed tendency of the hole orbital states is also consistent with the recent soft-x-ray-ARPES result
suggesting that the holes are expected to be nearly equally doped into the two $e_{\mathrm{g}}$ orbitals at $x=1.1$. \cite{NiFS2}
The existence of nominal Ni$^{4+}$ site above $x=1.0$ appears
to suppress the CB-type ordering or correlation and then drive the succeeding insulator-metal transition.
In addition, in accord with the change of the orbital sector, lattice constants as shown in Figs. 3(b) and 3(c) exhibit the concomitant nonmonotonic dependence.
With sharp increase in density of the Ni$^{3+}$ site with the $3z^2-r^2$-orbital hole,
the $a$- and $c$-axis lattice parameters start to increase and decrease above $x\sim 0.5$, respectively,
reflecting the strong electron-lattice coupling characteristic of the $e_{\mathrm{g}}$ orbital systems.
These experimental results indeed support that the reported high-energy pseudogap is ascribable
to the short-range or dynamic CB-type charge correlation persisting due to the orbital degree of freedom.

\begin{figure}
\begin{center}
\includegraphics*[width=5.6cm]{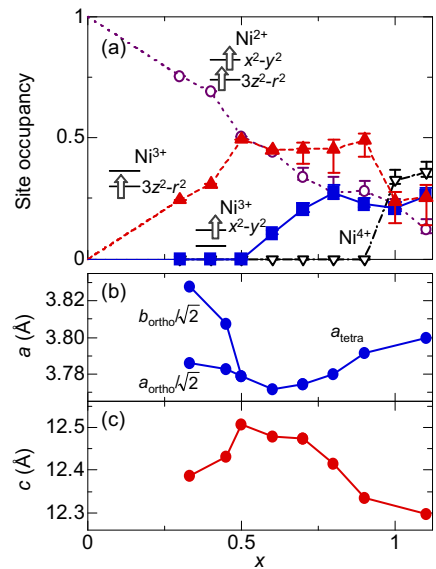}
\caption{(Color online).
(a) Doping variation of the site occupancy with the respective valence and orbital states for the hole doping in Nd$_{2-x}$Sr$_x$NiO$_4$,
as estimated by fitting and quantifying the peaks B-D in the O $K$-pre-edge spectra for {\EparaC}.
Room-temperature lattice constants along (b) $a$- and (c) $c$-axis directions for Nd$_{2-x}$Sr$_x$NiO$_4$ single crystals.
The crystal structure is low-temperature orthorhombic for $x\leq0.45$ and high-temperature tetragonal for $x\geq0.50$, respectively.
}
\label{fig3}
\end{center}
\end{figure}

Figure 4(a) gives the doping variation of the in-plane resistivity $\rho _{ab}$.
$\rho _{ab}$ gradually decreases with further doping holes into the CB-type charge ordered phase.
Above $x=1.0$ the conductivity $\sigma _{ab} (\equiv 1/\rho _{ab})$ shows a finite value $\sigma _{0}$ in the zero temperature limit,
which indicates that the metallic ground state emerges.
The low-temperature weak upturn therein can be fitted with the equation $\sigma _{ab} = \sigma _{0} + A\sqrt{T}$, \cite{NiFS1, Nithin}
suggesting three-dimensional weak localization. \cite{WL}
The pseudogap formation cannot be accounted for in terms of the so-called soft-gap, \cite{ES}
since the energy scale of such weak localization should be far smaller than that of the high-energy pseudogap,
while that might be an origin of the discrepancy between the conductivity and the extrapolated value in the optical conductivity spectra shown later.

\begin{figure}
\begin{center}
\includegraphics*[width=8.6cm]{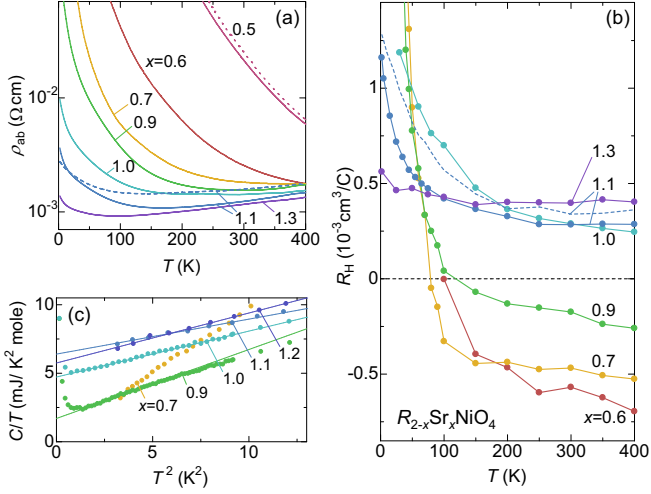}
\caption{(Color online).
Temperature and doping variations of the in-plane (a) resistivity $\rho _{ab}$ and (b) Hall coefficient $R _{\mathrm{H}}$ in $R_{2-x}$Sr$_x$NiO$_4$.
Dotted, solid, and broken lines show the results obtained for $R = \mathrm{La}$, Nd, and Eu, respectively.
Little $R$ dependence is observed as also confirmed in the previous reports. \cite{NiFS1, Nisigma1}
(c) Doping dependence of the low-temperature specific heat in a form of $C/T$ vs. $T^2$ plot for $R = \mathrm{Eu}$.
Straight lines represent the extrapolation toward zero temperature to estimate the electronic specific heat coefficient ($\gamma$, Fig. 6(c)).
The deviation from the lines below $T\sim1$ K arises from the nuclear Schottky component.
In addition, the curved shape with the steeper slope at $x=0.7$ is due to an additional low-temperature entropy component appearing around the low-doping region. \cite{Nigamma}
}
\label{fig4}
\end{center}
\end{figure}

In Fig. 4(b) we show the temperature and doping dependence of the in-plane Hall coefficient $R _{\mathrm{H}}$.
In the metallic region at $x=1.0$-1.3,
positive values nearly saturated in the high-temperature limit ($\sim 4\times 10^{-4}$ cm$^3$/C)
correspond to the case of one hole carrier per Ni site, indicating a typical metal with the large Fermi surface.
Namely, the ($\pi, \pi$)-centered large hole surface (Fig. 1(b)) is interpreted as dominantly contributing
to $R _{\mathrm{H}}$ and its temperature variation, regardless of the existence of the electron pocket. \cite{NiFS2}
For $x=1.0$ and 1.1, $R _{\mathrm{H}}$ shows monotonic increase with lowering temperature,
while it becomes nearly-temperature independent at $x=1.3$.
As discussed later, the pseudogap formation is a plausible origin of this anomalous metallic behavior.
Below $x=0.9$, on the other hand, the sign of $R _{\mathrm{H}}$ becomes negative in the high-temperature region.
This doping-induced sign change results from the expansion (shrink) of the observed electron (hole) Fermi surfaces with increasing the electron number.
In other words, the sign change of $R _{\mathrm{H}}$ in the high-temperature metallic state also confirms the multi-band/orbital nature as revealed by the XAS study.
The low-temperature rapid divergence with sign inversion is therefore ascribable to 
topology change of the predominant Fermi surface from an electron surface to a small hole pocket
through the antiferromagnetic spin ordering ($T_{\mathrm{N}}\sim80$ K). \cite{Nimsr}

Figure 4(c) presents the doping variation of the low-temperature specific heat
in the layered nickelates with no magnetic moment on the $R=\mathrm{Eu}$ site.
In the $C/T$ vs. $T^2$ plot, the electronic specific-heat coefficient $\gamma$
can be estimated as an intersection of the extrapolation toward zero temperature.
Above $x=1.0$, $\gamma$ certainly has a finite value as a sign of the metallic ground state.
A noteworthy feature is that its magnitude remains small ($\sim 6$ mJ/K$^2$ mole) and
shows no enhancement in the vicinity of the metal-insulator transition.
This behavior is in contrast with the critical mass renormalization effect,
which is as intrinsically observed in typical filling-control Mott transition systems
such as La$_{1-x}$Sr$_x$TiO$_3$ \cite{gammaLSTO} and La$_{1-x}$Sr$_x$VO$_3$\cite{gammaLSVO}.
The result suggests that the metal-insulator transition in the layered nickelate is driven
not by the carrier mass divergence but by the carrier number vanishment.
This situation is analogous to the superconducting layered cuprate case as studied in La$_{2-x}$Sr$_x$CuO$_4$,
in which the estimated $\gamma$ value in the normal state shows no critical enhancement
($\sim 9$ mJ/K$^2$ mole) \cite{gammaCu1, gammaCu2}
as a hallmark of the carrier-vanishing Mott transition under the presence of the pseudogap. \cite{MIT, PG}
 
\begin{figure}
\begin{center}
\includegraphics*[width=8cm]{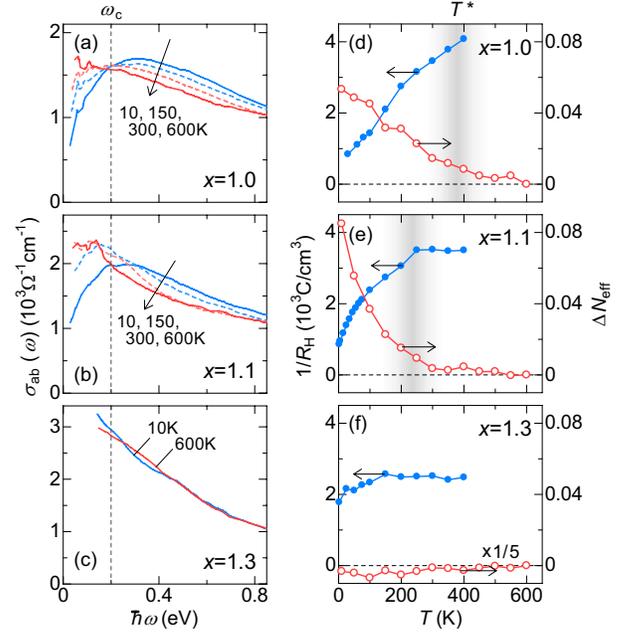}
\caption{(Color online).
(a)-(c) Temperature variation of the in-plane optical conductivity spectra $\sigma _{ab} (\omega)$ in Nd$_{2-x}$Sr$_x$NiO$_4$ with $x=1.0$, 1.1, and 1.3.
(d)-(f) Temperature dependence of the inverse of the Hall coefficient $1/R _{\mathrm{H}}$ and
the low-energy spectral weight (effective number of electrons $\Delta N_{\mathrm{eff}}$ at $\omega_{\mathrm{c}}=0.2$ eV, see text for details)
for the corresponding doping levels.
Remarkable decrease and increase of $1/R _{\mathrm{H}}$ and $\Delta N_{\mathrm{eff}}$
are observed below the characteristic temperature $T^*$ (indicated by vertical shadows), respectively.
}
\label{fig5}
\end{center}
\end{figure}

Now let us move on to the detailed charge dynamics in the metallic region adjacent to the insulating phase.
Figures 5(a)-(c) show the in-plane optical conductivity spectra $\sigma _{ab} (\omega)$
ranging from the anomalous metallic state ($x=1.0$-1.1) to the normal metal ($x=1.3$).
We first focus on the ground state at 10 K.
With increasing $x$, a mid-infrared peak with the real-gap structure ($\sim 0.6$ eV at $x=0.5$) \cite{Nisigma1, NisigmaCB, NioptCB}
gradually evolves into the pseudogap structure at $x=1.0$-1.1
and then turns to the normal metallic spectrum with increasing conductivity toward zero energy at $x=1.3$.
This doping variation can be explained in terms of the CB-type charge correlation persisting far beyond $x=0.5$.
The pseudogap magnitude ($\sim 0.2$ eV) provides a scale of the binding energy of the CB-type-correlated holes,
as determined by the intersite Coulomb and breathing-type Jahn-Teller interactions.
As suggested for the case of the layered manganites, \cite{Mngap}
softening of such a breathing phonon mode can potentially drive the pseudogap formation through the metal-insulator transition.
With increasing $x$ above $x=0.5$, the gap magnitude decreases reflecting the gradual weakening of the charge-order amplitude,
and the remnant polaronic behavior may give rise to the incoherent carrier dynamics in the anomalous metallic region.
While little temperature dependence is observed for the Drude-like response at $x=1.3$,
the pseudogap structure at $x=1.0$-1.1 easily collapses into a broadened gapless feature with increasing temperature,
and shows saturation of the temperature dependence at several hundred kelvin.
Such a spectral variation indicates that the CB-type charge correlation develops below some characteristic temperature in the pseudogap regime.

To analyze the temperature-dependent pseudogap,
we show in Figs. 5(d)-(f) the temperature variation of the spectral weight below the pseudogap energy,
namely the effective number of electrons $\Delta N_{\mathrm{eff}}$ at 0.2 eV.
{\Neff} is defined by the following relation,
\begin{equation*}
\begin{split}
N_{\mathrm{eff}}=\frac{2m_{\mathrm{0}}V}{\pi e^2}\int_0^{\omega_{\mathrm{c}}}\sigma (\omega) d\omega .
\end{split}
\end{equation*}
Here $m_{\mathrm{0}}$, $V$, and $\omega_{\mathrm{c}}$ ($=0.2$ eV) are
the free-electron mass, the cell volume containing one Ni atom, and the cut-off energy
determined from the isosbestic point energy of $\sigma _{ab} (\omega)$.
$\Delta N_{\mathrm{eff}}$ at a given temperature $T$ is
defined as $\Delta N_{\mathrm{eff}} = N_{\mathrm{eff}} (600\: \mathrm{K}) -N_{\mathrm{eff}} (T) $,
representing the amount of the spectral weight loss below 0.2 eV.
For $x=1.0$ and 1.1, the $\Delta N_{\mathrm{eff}}$ value starts to notably increase below $T^* \sim 400$ and 250 K, respectively.
The characteristic temperature $T^*$ can be interpreted as an onset temperature of the pseudogap formation.
Together with $\Delta N_{\mathrm{eff}}$, we show the Hall coefficient in a form of $1/R _{\mathrm{H}}$,
which also provides a measure for the carrier number in the present case of the nearly single band.
Below $T^*$, $1/R _{\mathrm{H}}$ at $x=1.0$-1.1 shows remarkable decrease
from the high-temperature value corresponding to the large hole surface volume without the pseudogap.
At $x=1.3$, on the other hand, the decrease is hardly observed.
A plausible origin of the temperature variation is the partial disappearance of the Fermi surface or
the decrease of the mobile carrier number due to the CB-type charge correlation.
Our results indicate that in the pseudogap regime the coherent carrier dynamics is partly suppressed
in response to the strength of the evolving charge correlation.
In particular, its suppression is expected to selectively occur centered around the ($\pi, 0$) region (Fig.1 (b)),
mutually connected with the CB-type ($\pi, \pi$) wave vector.

\begin{figure}
\begin{center}
\includegraphics*[width=8.6cm]{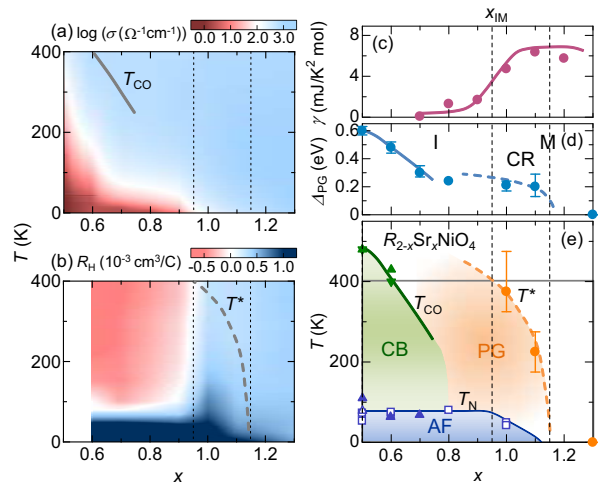}
\caption{(Color online).
Evolution of (a) the in-plane conductivity $\sigma _{ab}$ and (b) the Hall coefficient $R _{\mathrm{H}}$ in the $T$-$x$ diagram.
Doping variation of (c) the electronic specific-heat coefficient $\gamma$ and
(d) the pseudogap (PG) magnitude $\varDelta _{\mathrm{PG}}$ (see text for details).
They indicate the insulator (I)-metal (M) ground state boundary and the proximate critical doping region (CR),
roughly denoted by the vertical broken lines.
(e) Electronic phase diagram of $R_{2-x}$Sr$_x$NiO$_4$ around the metal-insulator transition.
The pseudogap formation temperature $T^*$ (see text for details) are presented together with
the checkerboard (CB)-type charge order and antiferromagnetic (AF) spin order transition temperatures
(upward triangles, \cite{Nineutron} downward triangles, \cite{Nisigma1} and squares \cite{Nimsr}).
The open and filled marks indicate the data for $R=\mathrm{La}$ and Nd, respectively;
in this system the critical temperatures of the spin and charge orderings have been confirmed to be
least influenced by the variation of the $R$ species. \cite{Nisigma1, Nineutron, Niorthostripe}
The solid and broken curves are merely the guide to the eyes.
}
\label{fig6}
\end{center}
\end{figure}

In Fig. 6 we summarize temperature and doping variations of the various physical quantities
characterizing the metal-insulator transition in the layered nickelates.
Figures 6(a) and 6(b) give the evolution of $\sigma _{ab}$ and $R _{\mathrm{H}}$ in the $T$-$x$ phase diagram, respectively.
The $\sigma _{ab}$ downturn temperature systematically decreases along with $T_{\mathrm{CO}}$,
and the metallic ground state emerges above $x=1.0$ as confirmed by the finite value of $\sigma _{ab}$ at the lowest temperature.
As indicated in the sign inversion of the high-temperature $R _{\mathrm{H}}$ value,
the dominant Fermi surface changes between $x=0.9$ and 1.0
from the $\Gamma$-centered electron one to the ($\pi, \pi$)-centered hole one.
In the anomalous metallic state ($x=1.0$-1.1), $R _{\mathrm{H}}$ shows large increase below $T^*$, 
while the nearly temperature-independent behavior is observed in the normal metal ($x=1.3$).
In addition, $\gamma$ in the metallic region is relatively small and shows no critical enhancement
at the boundary of the metal-insulator transition, as seen in Fig. 6(c).
These findings again demonstrate that the metal-insulator transition in this nickelate system is driven
not by the carrier mass divergence but by the carrier number vanishment due to the pseudogap formation.
In Figs. 6(d) and 6(e) we show the doping variation of the pseudogap magnitude $\varDelta _{\mathrm{PG}}$
estimated from the isosbestic point of $\sigma _{ab} (\omega)$,
together with $T^*$, $T_{\mathrm{CO}}$, and $T_{\mathrm{N}}$.
With increasing the hole doping level above $x=0.5$,
$\varDelta _{\mathrm{PG}}$ and $T_{\mathrm{CO}}$ gradually decrease in contrast to $T_{\mathrm{N}}$,
reflecting the weakening CB-type charge correlation.
Even in the critical region at $x=1.0$-1.1, however,
the pseudogap structure with $\varDelta _{\mathrm{PG}}$ of about 0.2 eV develops with lowering temperature below $T^*$,
showing similar temperature-doping dependence as in the layered cuprates.
Combined with the former ARPES study, \cite{NiFS1}
it is likely that below $T^*$ the coherent carrier dynamics around the ($\pi, 0$) region is strongly and momentum (\textbf{k})-selectively suppressed
with evolution of the CB-type charge correlation.
In the normal metal at $x=1.3$, the coherent state is expected to appear all over the momentum region of the large hole surface.

\section{Conclusion}

In conclusion, first we have reported on the detailed doping variation of the hole orbital states in the layered nickelate,
and confirmed the possibility that the CB-type charge correlation can persist up to the critical doping region because of the multi-orbital nature.
Then we have systematically investigated the pseudogap dynamics and concomitant critical behavior in the metal-insulator transition.
While no critical mass enhancement is observed at the boundary of the metal-insulator transition,
the carrier number markedly decreases in accord with clear evolution of the pseudogap structure.
Our results indicate that the CB-type charge correlation indeed dominates the charge dynamics in the pseudogap regime
and that the pseudogap evolution induces the metal-insulator transition
by gradually and momentum (\textbf{k})-selectively suppressing the density of states or the quasiparticle spectral weight at the Fermi level.
The present findings will provide better understanding of charge dynamics near the metal-charge-ordered-insulator transition
as well as of the \textbf{k}-dependent pseudogap formation in a broader class of systems.

\begin{acknowledgments}
We would like to thank H. Wadati for enlightening discussions.
This work was partly supported by a Grant-in-Aid for Scientific Research (Grants No. 20340086 and No. 21224008) from JSPS,
and by Funding Program for World-Leading Innovative R\&D on Science and Technology (FIRST Program).
M.U. acknowledges support by a Grant-in-Aid for the JSPS Fellowship program (Grant No. 21-5941).
\end{acknowledgments}

\end{document}